\documentclass{article}
\usepackage{amsfonts}
\usepackage{amsmath}
\usepackage{amssymb}
\usepackage{graphicx}
\usepackage{harvard}
\begin{document}

\title{Nonequilibrium sum rules for the Holstein model}
\author{J.~K.~Freericks$^1$,\\
Khadijeh Najafi$^1$\\
A.~F.~Kemper$^2$,\\ and T.~P.~Devereaux$^{3,4}$\\ \\
$^1$Department of Physics, Georgetown University,\\
37th and O Sts. NW, Washington, DC 20057, USA\\
$^2$Lawrence Berkeley National Laboratory,\\
1 Cyclotron Road, Berkeley, CA 94720, USA\\
$^3$Stanford Institute for Materials and Energy Science,\\
SLAC National Accelerator Laboratory,\\
Menlo Park, CA 94025, USA\\
$^4$Geballe Laboratory for Advanced Materials,\\
Stanford University, Stanford, CA 94305, USA}

\date{}
\maketitle

\begin{abstract}

We derive frequency moment sum rules for the retarded electronic  Green's function and self-energy of the Holstein model for both 
equilibrium and nonequilibrium cases. We also derive sum rules for the phonon propagator in equilibrium and nonequilibrium. These sum rules allow one to benchmark nonequilibrium calculations and
help with interpreting the behavior of electrons driven out of equilibrium by an applied electric field. We exactly evaluate the
sum rules when the system is in the atomic limit. We also discuss the application of these sum rules to pump/probe experiments like time-resolved angle-resolved photoemission spectroscopy.


\end{abstract}

\section{Introduction}

In recent years, we have seen significant advances in time-resolved experiments on systems that have strong
electron-phonon interactions~\cite{berkeley,bovensiepen}. These experiments study how energy is transferred between the electronic and phononic parts of the system. One of the interesting effects that has been seen in these experiments
is the so-called phonon-window effect~\cite{prx}, where electrons with energies farther than the phonon frequency from the Fermi level relax quickly back to equilibrium after the pulsed field is applied, but those close to the Fermi level relax on a much longer time scale, because their relaxation involves multiparticle processes due to a restricted phase space. It is clear that this experimental and theoretical work is just starting to analyze electron-phonon interacting systems in the time domain. Hence, any exact results that can be brought to bear on this problem will be important. 

In this work, we derive sum rules for the zeroth and first two moments of the retarded electronic Green's function and for the zeroth moment of the retarded self-energy. The moment sum rules have already been derived in equilibrium~\cite{kornilovitch,rosch}, but they actually hold true, unchanged, in nonequilibrium as well~\cite{sumrules1,sumrules2,sumrules3,sumrules4}. With these sum rules, one can understand how the electron-phonon interaction responds to nonequilibrium driving, and how different response functions will behave.

We start with the so-called Holstein model~\cite{holstein1,holstein2}, given by the following Hamiltonian in the Schroedinger representation:
\begin{equation}
\mathcal{H}(t)=-\sum_{ij\sigma}t_{ij}(t)c^\dagger_{i\sigma}c^{}_{j\sigma}+\sum_{i\sigma}[g(t) x_i-\mu] c^\dagger_{i\sigma}c^{}_{i\sigma}+\sum_i\frac{p_i^2}{2m}+\frac{1}{2}\kappa\sum_ix_i^2
\label{eq: ham}
\end{equation}
where $c^\dagger_{i\sigma}$ ($c^{}_{i\sigma}$) are the fermionic creation (annihilation) operators for an electron at lattice site $i$ with spin $\sigma$ (with anticommutator $\{c^{}_{i\sigma},c^\dagger_{j\sigma\prime}\}_+=\delta_{ij}\delta_{\sigma\sigma'}$), and $x_i$ and $p_i$ are the phonon coordinate and momentum (with commutator $[x_i,p_j]_-=i\hbar\delta_{ij}$), respectively. The hopping $-t_{ij}(t)$ between lattice sites $i$ and $j$ can be time dependent [for example, an applied electric field corresponds to the Peierls' substitution~\cite{peierls}], $\mu$ is the chemical potential for the electrons, $g(t)$ is the time-dependent electron-phonon interaction, $m$ is the mass of the optical (Einstein) phonon and $\kappa$ is the corresponding spring constant. The frequency of the phonon is $\omega=\sqrt{\kappa/m}$. It is often convenient to also express the phonon degree of freedom in terms of the raising and lowering operators $a^\dagger_i$ and $a^{}_i$ (with commutator $[a^{}_i,a^\dagger_j]_-=\delta_{ij}$) with $x_i=(a^\dagger_i+a^{}_i)\sqrt{\hbar/(2m\omega)}$ and $p_i=(-a^\dagger_i+a^{}_i)\sqrt{\hbar m\omega/2}/i$. This Hamiltonian involves electrons that can hop between different sites on a lattice and interact with harmonic Einstein phonons that have the same phonon frequency for every lattice site. The hopping and the electron-phonon coupling are taken to be time dependent for the nonequilibrium case. We set $\hbar=1$ and $k_B=1$ for the remainder of this work.

\section{Formalism for the electronic sum rules}

Since we will be working in nonequilibrium, we need to allow the time to lie somewhere on the Kadanoff-Baym-Keldysh contour, which runs in the positive time direction from $t_{min}$ to $t_{max}$, back to $t_{min}$ and down the imaginary axis to $t_{min}-i\beta$, with $\beta=1/T$ the inverse temperature (it is assumed the system is in equilibrium at time $t_{min}$). Allowing the time to be chosen anywhere on the contour, the contour-ordered electronic Green's function is defined by
\begin{equation}
G^c_{ij\sigma}(t,t')=-i{\rm Tr}\mathcal{T}_c e^{-\beta\mathcal{H}(t_{min})}c^{}_{i\sigma}(t)c^\dagger_{j\sigma}(t')/\mathcal{Z},
\label{eq: g_contour}
\end{equation}
where $\mathcal{T}_c$ denotes time ordering along the contour, and $\mathcal{Z}={\rm Tr}\exp[-\beta \mathcal{H}(t_{min})]$, with the system in equilibrium at the initial time $t_{min}$ at a temperature $T=1/\beta$. The Fermi operators are written in the Heisenberg representation $c_{i\sigma}(t)=U^\dagger(t,t_{min})c_{i\sigma}U(t,t_{min})$, with $U(t,t')$ the evolution operator from time $t'$ to time $t$. The evolution operator satisfies $idU(t,t')/dt=\mathcal{H}(t)U(t,t')$ and $U(t,t)=1$. From the contour-ordered Green's function, one can extract all of the needed Green's functions, like the so-called lesser Green's function and the retarded Green's function, which we consider in detail here, and which is defined by
\begin{equation}
G^R_{ij\sigma}(t,t')=-i\theta(t-t'){\rm Tr}e^{-\beta\mathcal{H}(t_{min})}\{c^{}_{i\sigma}(t),c^\dagger_{j\sigma}(t')\}_+/\mathcal{Z},
\label{eq: g_retarded}
\end{equation}
where $\{.,.\}_+$ denotes the anticommutator. Converting to Wigner's average and relative times $t_{ave}=(t+t')/2$ and $t_{rel}=t-t'$, we can find the frequency-dependent retarded Green's function for each average time via
\begin{equation}
G^R_{ij\sigma}(t_{ave},\omega)=\int_0^\infty dt_{rel} e^{i\omega t_{rel}}G^R_{ij\sigma}(t_{ave}+\frac{1}{2}t_{rel},t_{ave}-\frac{1}{2}t_{rel}).
\label{eq: gret_w}
\end{equation}
The $n$th spectral moment in real space is then defined via
\begin{equation}
\mu^{Rn}_{ij\sigma}(t_{ave})=-\frac{1}{\pi}\int_{-\infty}^{\infty}d\omega \,\omega^n{\rm Im}G^R_{ij\sigma}(t_{ave},\omega).
\label{eq: moment_def}
\end{equation}
The moments are more convenient to evaluate as derivatives in time
\begin{eqnarray}
&~&\mu^{Rn}_{ij\sigma}(t_{ave})=\\
&~&{\rm Im}\Biggr\{
\frac{i^{n+1}}{\mathcal{Z}}\frac{d^n}{dt_{rel}^n} {\rm Tr}e^{-\beta \mathcal{H}(t_{min})}
\{c^{}_{i\sigma}(t_{ave}+\frac12 t_{rel}),c^\dagger_{j\sigma}(t_{ave}-\frac12 t_{rel})\}_+\Biggr|_{t_{rel}=0^+}\Biggr\}.\nonumber
\label{eq: moment_deriv}
\end{eqnarray}
These time derivatives can be replaced by partial time derivatives with respect to time-dependent terms in the Hamiltonian plus commutators with the Hamiltonian.  In particular, we find that
\begin{equation}
\mu^{R0}_{ij\sigma}(t_{ave})={\rm Tr}e^{-\beta \mathcal{H}(t_{min})}
\{c^{}_{i\sigma}(t_{ave}),c^\dagger_{j\sigma}(t_{ave})\}_+/\mathcal{Z}
\end{equation}
for the zeroth moment,
\begin{eqnarray}
\mu^{R1}_{ij\sigma}(t_{ave})&=&-\frac12\langle \{[\mathcal{H}_H(t_{ave}),c^{}_{i\sigma}(t_{ave})]_-,c^\dagger_{j\sigma}(t_{ave})\}_+\rangle\nonumber\\
&+&\frac12\langle\{c^{}_{i\sigma}(t_{ave}),[\mathcal{H}_H(t_{ave}),c^\dagger_{j\sigma}(t_{ave})]_-\}_+\rangle,
\end{eqnarray}
for the first moment, where the angle brackets denote the trace over all states weighted by the density matrix ($\langle O\rangle={\rm Tr}\exp[-\beta\mathcal{H}(t_{min})]O/\mathcal{Z}$), the symbol $[.,.]_-$ denotes the commutator, and the subscript $H$ on the Hamiltonian indicates that it is in the Heisenberg representation. The second moment is more complicated and satisfies
\begin{eqnarray}
\mu^{R2}_{ij\sigma}(t_{ave})&=&-\frac14\langle\{[\mathcal{H}_H(t_{ave}),[\mathcal{H}_H(t_{ave}),c^{}_{i\sigma}(t_{ave})]_-]_-,c^\dagger_{j\sigma}(t_{ave})\}_+\rangle\nonumber\\
&-&\frac12\langle\{[\mathcal{H}_H(t_{ave}),c^{}_{i\sigma}(t_{ave})]_-,[\mathcal{H}_H(t_{ave}),c^\dagger_{j\sigma}(t_{ave})]_-\}_+\rangle\nonumber\\
&+&\frac14 \langle\{c^{}_{i\sigma}(t_{ave}),[\mathcal{H}_H(t_{ave}),[\mathcal{H}_H(t_{ave}),c^\dagger_{j\sigma}(t_{ave})]_-]_-\}_+\rangle\nonumber\\
&+&\frac14{\rm Im}\langle \{[\mathcal{H}^\prime_H(t_{ave}),c^{}_{i\sigma}(t_{ave})]_-,c^\dagger_{j\sigma}(t_{ave})\}_+\rangle\nonumber\\
&+&\frac14{\rm Im}\langle\{c^{}_{i\sigma}(t_{ave}),[\mathcal{H}^\prime_H(t_{ave}),c^\dagger_{j\sigma}(t_{ave})]_-\}_+\rangle,
\label{eq: second_formal}
\end{eqnarray}
where the prime indicates it is the Heisenberg representation of the time derivative of the  Schroedinger representation Hamiltonian [{\it i.~e.}, $\mathcal{H}_{H}^\prime(t_{ave})= $ \\
$U^\dagger(t_{ave},-\infty)\partial \mathcal{H}_S(t)/\partial t|_{t=t_{ave}} U(t_{ave},-\infty)$]. One can directly see that the two terms with the derivative of the Hamiltonian [last two lines of Eq.~(\ref{eq: second_formal})] are equal and opposite and hence cancel.

These moments can now be evaluated straightforwardly, although the higher the moment is the more work it takes. We find the well-known result
\begin{equation}
\mu^{R0}_{ij\sigma}(t_{ave})=\delta_{ij}
\end{equation}
for the zeroth moment.  The first moment satisfies
\begin{equation}
\mu^{R1}_{ij\sigma}(t_{ave})=-t_{ij}(t_{ave})-\mu\delta_{ij}+g(t_{ave})\langle x_i(t_{ave})\rangle\delta_{ij}
\end{equation}
and the second moment becomes
\begin{eqnarray}
\mu^{R2}_{ij\sigma}(t_{ave})&=&\sum_kt_{ik}(t_{ave})t_{kj}(t_{ave})+2\mu t_{ij}(t_{ave})+\mu^2\delta_{ij}\\
&-&t_{ij}(t_{ave})g(t_{ave})\langle x_i(t_{ave})+x_j(t_{ave})\rangle-2\mu g(t_{ave})\langle x_i(t_{ave})\rangle\delta_{ij}\nonumber\\
&+&g^2(t_{ave})\langle x_i^2(t_{ave})\rangle\delta_{ij}.
\nonumber
\end{eqnarray}
Unlike in the case of the Hubbard or Falicov-Kimball model, where the sum rules relate to constants or simple
expectation values~\cite{sumrules1,sumrules2,sumrules3}, one can see here that one needs to know things like the average phonon coordinate and its fluctuations in order to find the moments.  We will discuss this further below.

Our next step, is to calculate the self-energy moments, which are defined via
\begin{equation}
C^{Rn}_{ij\sigma}(t_{ave})=-\frac{1}{\pi}\int d\omega \,\omega^n{\rm Im}\Sigma^R_{ij\sigma}(t_{ave},\omega).
\end{equation} 
Note that the self-energy is defined via the Dyson equation
\begin{equation}
G_{ij\sigma}^R(t,t')=G_{ij\sigma}^{R0}(t,t')+\sum_{kl}\int d\bar t\int d\bar t' G^{R0}_{ik\sigma}(t,\bar t)\Sigma^R_{kl\sigma}(\bar t,\bar t')G^{R}_{lj\sigma}(\bar t',t'),
\end{equation}
where $G^{R0}$ is the noninteracting Green's function and the time integrals run from $-\infty$ to $\infty$.
The strategy for evaluating the self-energy moments is rather simple.  First, one writes the Green's function and self-energy in terms of the respective spectral functions
\begin{equation}
G^R_{ij\sigma}(t_{ave},\omega)=-\frac1\pi\int\frac{{\rm Im} G^R_{ij\sigma}(t_{ave},\omega^\prime)}{\omega-\omega^\prime+i0^+}d\omega^\prime
\end{equation}
and
\begin{equation}
\Sigma^R_{ij\sigma}(t_{ave},\omega)=\Sigma^R_{ij\sigma}(t_{ave},\infty)-\frac1\pi\int \frac{{\rm Im}\Sigma^R_{ij\sigma}(t_{ave},\omega^\prime)}{\omega-\omega^\prime+i0^+}d\omega^\prime.
\end{equation}
Next, one substitutes those spectral representations into the Dyson equation that relates the Green's function and self-energy to the noninteracting Green's function. By expanding all functions in a series in $1/\omega$ for large $\omega$, one finds the spectral formulas involve summations over the moments. By employing the exact values for the Green's function moments, one can extract the moments for the self-energy.  Details for the formulas appear elsewhere~\cite{sumrules2}. The end result is
\begin{equation}
\Sigma^R_{ij\sigma}(t_{ave},\infty)=g(t_{ave})\langle x_i(t_{ave})\rangle \delta_{ij}
\end{equation}
and
\begin{equation}
C^{R0}_{ij\sigma}(t_{ave})=g^2(t_{ave})[
\langle x_i^2(t_{ave})\rangle -\langle x_i(t_{ave})\rangle^2].
\end{equation}
So, the total strength (integrated weight) of the self-energy depends on the fluctuations of the phonon field.

\section{Formalism for the phononic sum rules}

The retarded phonon Green's function is defined in a similar way, via
\begin{equation}
D_{ij}^R(t,t')=-i\theta(t-t'){\rm Tr}e^{-\beta\mathcal{H}(t_{min})}[x_i(t),x_j(t')]_-/\mathcal{Z},
\end{equation}
with the operators in the Heisenberg representation.
The moments are defined in the same way as before.  First one converts to the average and relative time coordinates and
Fourier transforms with respect to the relative coordinate
\begin{equation}
D_{ij}^R(t_{ave},\omega)=\int dt_{rel}e^{i\omega t_{rel}}D_{ij}^R(t_{ave}+\frac12t_{rel},t_{ave}-\frac12 t_{rel}),
\label{eq: phonon}
\end{equation}
and then one computes the moments via
\begin{equation}
m^{Rn}_{ij}(t_{ave})=-\frac{1}{\pi}\int d\omega \,\omega^n {\rm Im} D_{ij}^R(t_{ave},\omega).
\end{equation}
The zeroth moment vanishes because $x_i$ commutes with itself at equal times. For the higher moments, we also derive a formula similar to what was used for the electronic Green's functions. In particular, we have
\begin{equation}
m^{R1}_{ij}(t_{ave})=-\frac12{\rm Im}\Big\{\langle [x^\prime_i(t_{ave}),x_j(t_{ave})]_-\rangle-
\langle [x_i(t_{ave}),x_j^\prime(t_{ave})]_-\rangle\Big\}
\end{equation}
for the first moment.  But $x^\prime_i(t_{ave})=-i[x_i(t_{tave}),\mathcal{H}_H(t_{ave})]_-=p_i(t_{ave})/m$, so we find
\begin{equation}
m^{R1}_{ij}(t_{ave})=\frac{1}{m}\delta_{ij}.
\end{equation}
Similarly,
\begin{eqnarray}
m^{R2}_{ij}(t_{ave})&=&-\frac14{\rm Im}i\Big\{\langle [x^{\prime\prime}_i(t_{ave}),x_j(t_{ave})]_-\rangle-2\langle[x^\prime_j(t_{ave}),x^\prime_j(t_{ave})]_-\rangle\nonumber\\
&+&\langle [x_i(t_{ave}),x^{\prime\prime}_j(t_{ave})]_-\rangle\Big\}.
\end{eqnarray}
Using the fact that $x^{\prime\prime}_i(t_{ave})=-i[p_i(t_{ave}),\mathcal{H}_H(t_{ave})]_-=-g(t_{ave})(n_{i\uparrow}(t_{ave})+n_{i\downarrow}(t_{ave}))-\kappa x_i(t_{ave})$, then shows that $m^{R2}_{ij}(t_{ave})=0$, since all commutators vanish. We don't analyze the phonon self-energy here. Unlike the electronic moments, the phononic moments, are much simpler, and do not require any expectation values to evaluate them.

We end this section by showing that the imaginary part of the retarded phonon Green's function is an odd function of $\omega$, which explains why all the even moments vanish.  If one evaluates the complex conjugate of the retarded phonon Green's function, one finds
\begin{equation}
D_{ij}^R(t,t')^*=i\theta(t-t'){\rm Tr}  [x_j(t'),x_i(t)]_-e^{-\beta \mathcal{H}(t_{min})}/\mathcal{Z}=D_{ij}^R(t,t')
\end{equation}
where the last identity follows by switching the order of the operators in the commutator and the invariance of the trace under a cyclic permutation. Hence, the phonon propagator in the time representation is real. Evaluating the frequency-dependent propagator, then shows that
$D_{ij}^{R*}(t_{ave},\omega)=D_{ij}^R(t_{ave},-\omega)$ by taking the complex conjugate of Eq.~(\ref{eq: phonon}). Hence the real part of the retarded phonon propagator in the frequency representation is an even function of frequency while the imaginary part is an odd function of frequency, and therefore all even moments vanish.

\section{Atomic limit of the Holstein model}

To get an idea of the phonon expectation values and the fluctuations, we solve explicitly for the expectation values for the Holstein model in the atomic limit, where $t_{ij}(t)=0$ and we can drop the site index from all operators. In this limit, one can exactly determine the Heisenberg representation operator $x(t)$ by solving the equation of motion for the Heisenberg representation operators $a(t)$ and $a^\dagger(t)$.  This yields
\begin{equation}
x(t)=\frac{ae^{-i\omega t}+a^\dagger e^{i\omega t}}{\sqrt{2m\omega}}-2{\rm Re}\left \{ie^{-\omega t}\int_0^tdt'e^{i\omega t'}g(t')\right \}\frac{n_\uparrow+n_\downarrow}{2m\omega}
\label{eq: x_heisenberg}
\end{equation}
where the electronic number operators commute with $\mathcal{H}$ now, so they have no time dependence. Since the atomic sites are decoupled from one another, we can focus on just a single site.  The partition function for a single site can be evaluated directly by employing standard raising and lowering operator identities. To begin, we note that the Hilbert space is composed of a direct product of the harmonic oscillator states 
\begin{equation}
|n\rangle=\frac{1}{\sqrt{n!}}\left ( a^\dagger \right )^n |0\rangle
\end{equation}
and the fermionic states
\begin{equation}
|0\rangle,\quad |\uparrow\rangle=c^\dagger_\uparrow|0\rangle,\quad |\downarrow\rangle=c^\dagger_\downarrow|0\rangle,\quad
|\uparrow\downarrow\rangle=c^\dagger_\uparrow c^\dagger_\downarrow |0\rangle.
\end{equation}
The partition function satisfies
\begin{equation}
\mathcal{Z}_{at}=\sum_{0,\uparrow,\downarrow,\uparrow\downarrow}\sum_{n^b=0}^\infty\langle n^b,n^f|\exp[-\beta \{
(gx-\mu)(n^f_\uparrow+n^f_\downarrow)+\omega (n^b+\frac12)\}]|n^b,n^f\rangle,
\end{equation}
where $n^f$ denotes the Fermi number operator and $n^b$ the Boson number operator (we will drop the $\exp[\beta\omega/2]$ term which provides just a constant). Since the product states are not eigenstates of $\mathcal{H}$, we cannot immediately evaluate the partition function.  Instead, we need to first go to the interaction representation with respect to the bosonic Hamiltonian in imaginary time (and we drop the constant term from the Hamiltonian), to find that
\begin{eqnarray}
\mathcal{Z}_{at}&=&{\rm Tr}_f{\rm Tr}_be^{-\beta\omega n^b}\mathcal{T}_{\tau} \exp\left [ -\int_0^\beta d\tau'\left \{\frac{e^{-\omega\tau'}a+e^{\omega\tau'}a^\dagger}{\sqrt{2m\omega}}g(t_{min}
)-\mu\right \}n^f\right ],\nonumber\\
&=&{\rm Tr}_f{\rm Tr}_b e^{\beta\omega n^b}U_I(\beta)
\label{eq: partition}
\end{eqnarray}
where the time-ordering operator is with respect to imaginary time and the time-ordered product is the evolution operator in the interaction representation and denoted by $U_I(\tau)$. Because the only operators that don't commute in the evolution operator are $a$ and $a^\dagger$, and their commutator is a c-number, one can get an exact representation for the evolution operator via the Magnus expansion~\cite{magnus}, as worked out in the Landau and Lifshitz~\cite{landau} or Gottfried~\cite{gottfried} texts. The end result for the time-ordered product in Eq.~(\ref{eq: partition}) becomes
\begin{eqnarray}
U_I(\beta)&=&\exp\left [ -\frac{g(t_{min})n^f}{\sqrt{2m\omega^3}}\left (1-e^{-\beta\omega}\right )a\right ]
\exp\left [ -\frac{g(t_{min})n^f}{\sqrt{2m\omega^3}}\left (e^{\beta\omega}-1\right )a^\dagger\right ]\nonumber\\
&\times&\exp\left [-\frac{g^2(t_{min})n^{f2}}{\sqrt{2m\omega^3}}\left ( e^{\beta\omega}-1-\beta\omega+\mu n^f\right )\right ],
\end{eqnarray}
which used the Campbell-Baker-Hausdorff theorem 
\begin{equation}
e^{A+B}=e^Be^Ae^{\frac12[A,B]_-}
\end{equation}
for the case when the commutator $[A,B]_-$ is a number, not an operator, to get the final expression.
We substitute this result for the evolution operator into the trace over the bosonic states, expand the exponentials of the $a$ and $a^\dagger$ operators in a power series, and evaluate the bosonic expectation value to find
\begin{eqnarray}
\mathcal{Z}_{at}&=&{\rm Tr}_f\sum_{m=0}^\infty\sum_{n=0}^\infty \frac{(n+m)!}{n!m!m!}\left (\frac{g^2(t_{min})n^{f2}}{2m\omega^3}\left (e^{\beta\omega}-1+e^{-\beta\omega}-1\right )\right )^m\nonumber\\
&\times&\exp\left [-\beta\omega n+\beta\mu n^f-\frac{g^2(t_{min})n^{f2}}{2m\omega^3}\left ( e^{\beta\omega}-1-\beta\omega\right )\right ].
\end{eqnarray}
Next, we use Newton's generalized binomial theorem
\begin{equation}
\sum_{n=0}^\infty\frac{(n+m)!}{n!m!}z^n=\frac{1}{(1-z)^{m+1}},
\end{equation}
to simplify the expression for the partition function to
\begin{equation}
\mathcal{Z}_{at}={\rm Tr}_f\exp\left [\frac{\beta g^2(t_{min})n^{f2}}{2m\omega^2}+\beta\mu n^f\right ]\frac{1}{1-e^{-\beta\omega}}.
\end{equation}
Performing the trace over the fermionic states then yields
\begin{equation}
\mathcal{Z}_{at}=\frac{1}{1-e^{-\beta\omega}}\left \{1+2e^{\beta\mu}\exp\left [\frac{\beta g^2(t_{min})}{2m\omega^2}\right ]
+e^{2\beta\mu}\exp\left [\frac{2\beta g^2(t_{min})}{m\omega^2}\right ]\right \}.
\end{equation}

We calculate expectation values following the same procedure, but inserting the relevant operators in the Heisenberg representation at the appropriate place, and carrying out the remainder of the derivation as done for the partition function.  For example, since the Fermi number
operators commute with the atomic Hamiltonian, they are the same operator in the Heisenberg and Schroedinger representations, and we immediately find that the electron density is a constant in time and is given by
\begin{equation}
\langle n_\uparrow+n_\downarrow\rangle=\frac{2e^{\beta\mu}\exp\left [\frac{\beta g^2(t_{min})}{2m\omega^2}\right ]
+2e^{2\beta\mu}\exp\left [\frac{2\beta g^2(t_{min})}{m\omega^2}\right ]}{1+2e^{\beta\mu}\exp\left [\frac{\beta g^2(t_{min})}{2m\omega^2}\right ]+e^{2\beta\mu}\exp\left [\frac{2\beta g^2(t_{min})}{m\omega^2}\right ]}.
\end{equation}

We next want to calculate $\langle x(t)\rangle$ and $\langle x^2(t)\rangle-\langle x(t)\rangle^2$, where the operator is in the Heisenberg representation, and given in Eq.~(\ref{eq: x_heisenberg}).  
It is straightforward but tedious to calculate the averages. After much algebra, we find
\begin{equation}
\langle x(t)\rangle=\langle n_\uparrow+n_\downarrow\rangle\left (-\frac{g(t_{min})}{m\omega^2}\cos\omega t-{\rm Re}\left \{ie^{-i\omega t}\int_0^tdt'e^{i\omega t'}\frac{g(t')}{m\omega}\right \}\right ).
\end{equation}
(Note that if we are in equilibrium, so $g(t)=g$ is a constant, then one finds $\langle x(t)\rangle=-\langle n_\uparrow+n_\downarrow\rangle g/m\omega^2$, which has no time dependence, as expected.) The fluctuation satisfies
\begin{eqnarray}
\langle x^2(t)\rangle-\langle x(t)\rangle^2&=&[\langle n_\uparrow\rangle(1-\langle n_\uparrow\rangle)+\langle n_\downarrow\rangle(1-\langle n_\downarrow\rangle)+2\langle n_\uparrow n_\downarrow\rangle-2\langle n_\uparrow\rangle\langle n_\downarrow\rangle ]\nonumber\\
&\times&\left [\frac{g(t_{min})}{m\omega^2}\cos\omega t+{\rm Re}\left \{ie^{-i\omega t}\int_0^tdt'e^{i\omega t'}\frac{g(t')}{m\omega}\right \}\right ]^2\nonumber\\
&+&\frac{1}{2m\omega}{\rm coth}\left (\frac{\beta\omega}{2}\right ),
\end{eqnarray}
which consists of two terms: a time-dependent piece (which becomes a constant when $g$ is a constant) that represents the quantum fluctuations due to the electron-phonon interaction and a phonon piece that varies with temperature (and is independent of $g$). The latter piece becomes large when $T\rightarrow\infty$ (being proportional to $T$ at high $T$), which tells us that fluctuations generically grow with increasing the temperature of the system, so that one expects the zeroth moment of the self energy to increase as the temperature increases, or if the system is heated up by being driven by a large electric field. Note that if one expands the self-energy perturbatively, as in Migdal-Eliashberg theory, then only the term independent of $g$ survives, as the other term is higher order in $g$ and lies outside of the Migdal-Eliashberg result~\cite{lex}.

\section{Discussion and applications of the sum rules}

One of the most important recent experiments in electron-phonon interacting systems involves time-resolved angle-resolved photoemission (tr-ARPES), which can be analyzed in such a way that one can extract information about the electronic self-energy~\cite{prx,lex}. If one assumes that the phonons form an infinite heat capacity bath, then they are not changed by the excitation of the electrons, and the fluctuations of the phonon field remain a constant as a function of time.  This leads to a self-energy that can transiently change shape as a function of time, but does not change its spectral weight. Recent calculations show precisely this behavior~\cite{prx,lex}. One can also understand it from the perturbation theory expansion, where a direct evaluation of the diagrams for the self-energy, and the sum rules for the electronic Green's functions, establish that the zeroth moment of the retarded electronic self-energy is a constant~\cite{lex}. What is perhaps more interesting, is when one treats a fully self-consistent system where the electrons and phonons both can exchange energy with one another, and the phonon bath properties change transiently. In this case, one has to examine the self-consistency for both the electrons and the phonons within the perturbation theory, and the general form of the sum-rules hold.  The calculations shown above in the atomic limit indicate that it is likely that adding energy into the phonon system increases the phonon fluctuations and thereby creates a stronger electronic self-energy.  One would expect there to be oscillations of the spectral weight as well. It is also likely that these ideas can be incorporated into the quantitative analysis of experiments that we expect to see occur over the next few years.

\section{Conclusions and future work}

In this work, we have shown the simplest sum rules for electrons interacting with phonons.  These sum rules have been established in equilibrium for some time now, but our work shows that they directly
extend to nonequilibrium.  We also established new sum rules for the phonon propagator.  In general, these sum rules are complicated to use, because they require one to determine both the average phonon expectation value and its fluctuations, so they might find their most important application to numerics as benchmarking, assuming one can calculate the relevant expectation values with the numerical techniques employed to solve the problem. But they also allow us to examine the physical behavior we expect to see if we look at how the moments might change in time due to the effect of a transient light pump applied to the system.  For example, we expect that as energy is exchanged from electrons to phonons, the electron self-energy should increase its spectral weight, with the opposite occuring as the phonons transfer energy back to the electrons. This result is not one that could have been easily predicted without the sum rules.

In the future, there are a number of ways these sum rules can be extended.  One can examine more realistic models, like the Hubbard-Holstein model and find those sum rules.  One can look into the effects of anharmonicity on the sum rules, and finally, one can carry out the calculations to higher order, to examine more moments. We plan to work on a number of these problems in the future.

\section{Acknowledgments}

This work was supported by the Department of Energy,
Office of Basic Energy Sciences, Division of Materials
Sciences and Engineering (DMSE) under Contracts
No. DE-AC02-76SF00515 (Stanford/SIMES), No. DE-FG02-08ER46542 (Georgetown), and No. DE-SC0007091
(for the collaboration). J.K.F. was also supported by the McDevitt
bequest at Georgetown

\bibliographystyle{agsm}
\bibliography{freericks_feis}

\end{document}